\documentclass[12pt]{article}
\usepackage{latexsym}
\usepackage{amsmath}
\usepackage{amsfonts}
\usepackage{amssymb}

\newcommand{\mH}{{\mathcal H}}
\title{On the triviality of higher derivative theories}
\author{K. Andrzejewski\thanks{e-mail: k-andrzejewski@uni.lodz.pl},
J. Gonera, P. Machalski,\\
\small Department of Theoretical Physics and Computer Science, \\
\small University of \L\'od\'z,\\
\small Pomorska 149/153, 90-236 {\L}\'od\'z, Poland
\\ 
K. Bolonek-Laso\'n \\
\small Department of Statistical Methods,\\
\small University of  \L\'od\'z,\\
\small Rewolucji 1905r 41/43, 90-214 L\'od\'z, Poland
}
\date{}
\begin{document}
\maketitle
\begin{abstract}
The higher-derivative theories with degenerate frequencies exhibit BRST symmetry (O. Rivelles, Phys. Lett. {\bf B577} (2003), 147). In the present paper meaning of BRST
 invariance condition is analyzed. The BRST symmetry is related   to  nondiagonalizability of the Hamiltonian and it is shown that BRST condition singles out the subspace 
spanned by proper eigenvectors of the Hamiltonian.
\end{abstract}
\section{Introduction}
Theories described by Lagrangians containing higher derivatives of basic variables seem to play some role in physics. Originally they were proposed as a method
 for dealing with ultraviolet divergences \cite{b1};  this idea appeared to be quite successful in the case of gravity: Einstein action supplied by the terms containing
 higher powers of curvature  leads to renormalizable \cite{b2}  and asymptotically  free \cite{b3} theory. Other examples of higher derivative theories include theory
 of radiation reaction  \cite{b4}, field theory on noncommutative  space-time  \cite{b5}, anyons \cite{b6} or string theories with extrinsic curvature \cite{b7}.
\par
Once we get used to the idea that second or higher derivatives can enter the Lagrangian leading to reasonable  theory the question arises whether  there exists
 a viable quantum  version of such dynamics. Following conservative route, to answer this question  we look first for the Hamiltonian formalism.
 Then we encounter serious difficulty: the Hamiltonian is unbounded from below \cite{b8}. Moreover, this property is related to large volumes of phase space so 
it cannot disappear on the quantum level due to the uncertainty principle (like, for example, it is the case for hydrogen atom). In some situations this frustrating property
of Ostrogradski Hamiltonian causes no problem.  This  is the case for linear systems like the celebrated Pais-Uhlenbeck  (PU)  one \cite{b9}, because there are no 
transitions between the states of different energies. However, linearity  is an idealization and any perturbation may cause instability. Another possibility of making the problem
 harmless arises if the exists some additional integral of motion bounded from below; then one can  try to construct an alternative hamiltonian formalism with the new constant
of motion  playing the role of Hamiltonian. This is again the case for PU oscillator  which is a completely integrable system \cite{b10}. In general, however, nothing
 allows us to believe that there should exist integrals of motion functionally independent of the one generated by time translation invariance. Therefore, we cannot expect
that, in general, there exists an alternative  hamiltonian formalism with bounded Hamiltonian.
\par 
There is some way out of this situation. In the Ostrogradski formalism one has introduce additional auxiliary canonical variables in order to make the dimension of
 phase space coinciding with the number of initial data needed for obtaining unique solution of Lagrange equations.
Starting from Ostrogradski variables one can then perform a complex canonical transformation. The only condition one has to impose is that the reality properties of basic
 Lagrangian variables remain  unaffected; the auxiliary variables may become complex. Upon standard quantization their properties under (hermitean) conjugation
will, in general, change. This would change the spectral properties of the Hamiltonian; the eigenvalues may become positive. However, there is some price to be payed:
to obtain an agreement between  transformation properties  of classical and quantum auxiliary variables under conjugation one has to modify the metric in the space of states. 
The new metric is necessarily indefinite. This is easily seen by referring  to the correspondence principle: the classical energies can be negative while the expectation values of 
the Hamiltonian with positive spectrum become negative only provided there exist states with negative norm.
\par 
The above-sketched scheme is encountered in the quantization scheme  proposed by Hawking and Hertog \cite{b11}. Their starting point is the higher derivative theory
defined by the action which is positive definite in Euclidean region. Then the Euclidean path integral
\begin{equation}
\label{e1}
\int {\mathcal D}e^{-S_E[\phi]},
\end{equation}
makes sense.  Hawking and Hertog gave well-defined prescription allowing to compute the Euclidean time transition amplitudes in terms of the above path integral.
\par
The real-time counterpart of the Hawking-Hertog construction for PU oscillator has been carefully analysed in \cite{b12}. It appeared that, for certain range of parameter, 
the Hamiltonian has purely real positive point  spectrum; while the metric is in the space of states indefinite and one obtains the scenario sketched above.
However, it appeared also that the quantum theory is well-defined not only in some  
subrange of parameters for which the integral (\ref{e1}) makes sense.
In particular, if both frequencies of PU oscillator coincide, the Hamiltonian is no longer diagonalisable \cite{b12},\cite{b13}.
Additionally, all eigenvectors, expect the ones corresponding to the lowest eigenvalue, have vanishing norm. One can hardly ascribe any physical meaning to the theory with
such properties.
\par
In an interesting  paper \cite{b14} Rivelles proposed an alternative point of view on some higher derivative theories. He considered the case of equal frequencies (in the limit when 
the interaction is switched off) and added the Faddeev-Popov ghost term to the Lagrangian. The dynamics of initial fields remains unaffected but the theory  acquires BRST symmetry.
Imposing the condition of BRST invariance one can show that the so called quartet mechanism \cite{b15} operates leading to trivial theory. The argument presented in Ref. \cite{b14}
is very elegant but it is based on the apriori assumption of BRST invariance of physical states. Contrary to the case of gauge theories,  the status of this condition is here slightly unclear.
We shall analyse this problem in more detail and show how it is related to the nondiagonalizability  of the Hamiltonian. We will consider the free quatric case (i.e. PU oscillator) but we 
conjecture our results survive in more general context (see below). Moreover, due to the translation invariance we can Fourier transform the spatial variables and reduce the problem to 
(1+0)-dimensional case.
\par The Lagrangian considered in Ref. \cite{b14}, with the ghost term neglected, reads ($\phi^*=\phi$,$ b^*=b$):
\begin{equation}
\label{e2}
L=-b(\frac{d^2}{dt^2}+2m^2)\phi+\frac{1}{2}b^2=\dot b\dot \phi-2m^2b\phi+\frac{1}{2}b^2+\textrm{ total\, derivative},
\end{equation}
and yields
\begin{equation}
\label{e3}
H=\Pi_{b}\cdot \Pi_{\phi}+2m^2b\cdot \phi-\frac{1}{2}b^2 .
\end{equation}
To make the contact with the model considered in Ref. \cite{b12} we perform the canonical transformation
\begin{alignat}{2}
\label{e4}
\phi &=\frac{q_1}{2m}, & \qquad b &= m(q_1+2ip_2) ,\nonumber \\
\Pi_{\phi}&=m(-iq_2+2p_1),  & \qquad \Pi_b &=\frac{iq_2}{2m}  .
\end{alignat}
Eqs. (\ref{e4}) imply that $q_2$ and $p_2$ are purely imaginary. The Hamiltonian, when expressed in new variables, reads
\begin{equation}
\label{e5}
H=iq_2p_1+2m^2p_2^2+\frac{m^2}{2}q_1^2+\frac{1}{2}q_2^2.
\end{equation}
One can show \cite{b12} that on the quantum level $H$ becomes hermitean (while the expectation values of $q_2$ and $p_2$  -- purely imaginary)
only provided the metric of the space of states is indefinite.
\par
To find the spectrum of $H$ the creation and annihilation operators are introduced by putting 
\begin{alignat}{2}
\label{e6}
&q_1=\frac{1}{\sqrt[4]{8m^2}}(g+g^{\star}), \quad p_1=-i\sqrt[4]{\frac{m^2}{2}}\big(g-g^{\star}+\frac{1}{2}(d-d^{\star})\big), \nonumber \\
&q_2=\sqrt[4]{\frac{m^2}{2}}(d-d^{\star}), \quad p_2=\frac{-i}{\sqrt[4]{8m^2}}\big(d+d^{\star}+\frac{1}{2}(g+g^{\star}\big),
\end{alignat}
which yields 
\begin{equation}
\label{e7}
[g,g^{\star}]=1,\quad [d,d^{\star}]=-1,
\end{equation}
as well as 
\begin{equation}
\label{e8}
H=-\sqrt{2}m(2d^{\star}d+dg^{\star}+d^{\star}g-1).
\end{equation}
The space of states is spanned by the vectors 
\begin{equation}
\label{e9}
|n_1,n_2\rangle=\frac{1}{\sqrt{n_1!}}\frac{1}{\sqrt{n_2!}}(g^{\star})^{n_1}(d^{\star})^{n_2}|0,0\rangle,\quad d|0,0\rangle=g|0,0\rangle=0,
\end{equation}
obeying
\begin{equation}
\label{e10}
\langle n_1,n_2|n_1' ,n_2'\rangle=(-1)^{n_2}\delta_{n_1n_1'}\delta_{n_2n_2'},\quad H|0,0\rangle=m\sqrt 2|0,0\rangle.
\end{equation}
Let 
\begin{equation}
\label{e11}
N=g^*g-d^*d,
\end{equation}
be the number operator 
\begin{equation}
\label{e12}
N|n_1,n_2\rangle=(n_1+n_2)|n_1,n_2\rangle.
\end{equation}
The following important identity holds
\begin{equation}
\label{e13}
H-\sqrt{2}m(N+1)=\sqrt{2}m(d+g)(d^{\star}+g^{\star}).
\end{equation}
Consider the subspace $\mH_n$ spanned by the vectors $|n_1,n_2\rangle$ such that $n_1+n_2=n$. Then on $\mH_n$, one has (by virtue of eq. (\ref{e13}))
\begin{equation}
\label{e14}
\big(H-\sqrt{2}m(N+1)\big)^{n+1}=0.
\end{equation}
Therefore on $\mH_n$, $H$ acquires a Jordan cell form corresponding to the eigenvalue $\sqrt 2 m (n+1)$.
The canonical basis in $\mH_n$ is spanned by the vectors $(g^*+d^*)^k(g^*-d^*)^{n-k}|0,0\rangle$,  $k=0,1,\ldots,n$; in particular,
\begin{equation}
\label{e15}
|\psi_n\rangle=(g^*+d^*)^n|0,0\rangle,
\end{equation}
is the  eigenvector of $H$, $H|\psi_n\rangle=\sqrt2 m(n+1)|\psi_n\rangle$.
We have also $\langle\psi_n|\psi_n\rangle=\delta_{n0}$ so among the eigenvectors of $H$ only the vacuum has a nonvanishing norm.
\par We conclude that $H$  is not fully diagonalisable. Instead, it can be put in Jordan block form. The spectrum of $H$ is purely real, positive and discrete, 
$E_n=\sqrt2 m(n+1)$, $n=0,1,\ldots$;
the Jordan cell corresponding to $E_n$ is (n+1)-dimensional. The important point is that, as  noted above, the only eigenvector of $H$ with nonvanishing norm 
is the vacuum $|\psi_0\rangle=|0,0\rangle$.
\par
In the case of nondiagonalizable Hamiltonian the important question arises what is the physical subspace of our theory. Apart from obvious condition that the metric 
is nonnegative when restricted to physical subspace there should exist further constraints related to the question of proper definition of energy of the system. The natural
assumption is that there exist stationary states and any physical state may be represented as their combination. Formally, this means that the physical subspace is
 spanned by the eigenvectors of $H$. In our case, all eigenvectors, except  vacuum, have vanishing norm.  Therefore the only physical state is the vacuum.
\par
Consider now the BRST-symmetry formulation of our model \cite{b14}. The starting point is the BRST-extended Lagrangian
\begin{equation}
\label{e16}
L=\dot{b}\dot{\phi}-2m^2b\phi+\frac{1}{2}b^2\pm i(\dot{\overline{c}}\dot{c}-2m^2\overline{c}c),
\end{equation}
where $c,\bar c$ are fermionic ghost fields obeying $c^*=c$, ${\bar c}^*=\bar c$.
Due  to the form of $L$ the ghost fields are free so they do not  influence the original dynamics. The BRST symmetry reads
\begin{equation}
\label{e17}
\delta c=0, \quad \delta b=0, \quad \delta\phi=\mp i\delta\lambda c, \quad \delta\overline{c}=\delta\lambda b,
\end{equation} 
where $\delta\lambda$ is anticommuting real parameters, $\delta \lambda^*=\delta\lambda$.
To see what is happening let us work in Heisenberg picture. First, we write out the modified Hamiltonian
\begin{equation}
\label{e18}
H=\mp i\Pi_{\overline{c}} \Pi_c\pm 2m^2\overline{c} c+\Pi_b \Pi_{\phi}-\frac{1}{2}b^2+2m^2b \phi.
\end{equation}                                  
For the bosonic sector the field equations and canonical commutation rules yield
\begin{align}
\label{e19}
b(t)&=\sqrt{m}\big(\beta e^{-i\sqrt{2}mt}+\beta^{\star}e^{i\sqrt{2}mt}\big), \nonumber \\
\phi(t)&=\frac{1}{2\sqrt{2m^3}}\Big(\big(\alpha+imt\beta\big)e^{-i\sqrt{2}mt}+\big(\alpha^{\star}-imt\beta^{\star}\big)e^{i\sqrt{2}mt}\Big), 
\end{align}
with
\begin{equation}
\label{e20}
[\alpha,\beta^{\star}]=1, \quad [\beta,\alpha^{\star}]=1, \quad [\alpha,\alpha^{\star}]=\frac{1}{\sqrt{2}}, 
\end{equation}
and the remaining commutators vanishing.
As far as the fermionic sector is concerned one obtains
\begin{align}
\label{e21}
c(t)&=\frac{1}{2}\sqrt{\frac{\sqrt{2}}{m}}\Big(ue^{-i\sqrt{2}mt}+u^{\star}e^{i\sqrt{2}mt}\Big), \nonumber \\
\overline{c}(t)&=\pm\frac{i}{2}\sqrt{\frac{\sqrt{2}}{m}}\Big(ve^{-i\sqrt{2}mt}-v^{\star}e^{i\sqrt{2}mt\Big)},\\
&\{u,v^{\star}\}=1, \quad \{v,u^{\star}\}=1.\nonumber
\end{align}
The BRST and ghost charges
\begin{align}
\label{e22}
Q_B&=\mp(b\dot{c}-c\dot{b}), \nonumber \\
Q_C&=\mp i(\dot{\overline{c}}c-\overline{c}\dot{c}\mp 1),
\end{align}
take the form (up to numerical factors)
\begin{equation}
\label{e23}
Q_B=\beta^*u-\beta u^*,\quad Q_c=v^*u-u^*v.
\end{equation}
To recover the original dynamics we have to get rid of ghosts. To this end we impose first the condition
\begin{equation}
\label{e24}
Q_C|\textrm{phys}\rangle=0 .
\end{equation}
There are four fermionic stats:  $|0\rangle_f$ \Big($u|0\rangle_f=0=v|0\rangle_f$\Big), $|u\rangle_f\equiv u^{\star}|0\rangle_f$, 
$|v\rangle_f\equiv v^{\star}|0\rangle_f$ and$|uv\rangle_f\equiv u^{\star}v^{\star}|0\rangle_f$. 
Eq.  (\ref{e24})  immediately yields
\begin{equation}
\label{e25}
|\textrm{phys}\rangle=|\textrm{phys}\rangle_b\otimes |0\rangle_f.
\end{equation}
Now, one demands the BRST condition to be fulfilled 
\begin{equation}
\label{e26}
Q_B|\textrm{phys}\rangle=0.
\end{equation}
Contrary to the case of theories invariant under some gauge group there exists here no apriori reason to impose (\ref{e26}).
However, once we do this the following constraint emerges
\begin{equation}
\label{e27}
\beta|\textrm{phys}\rangle_b=0.
\end{equation}
To find the meaning of this condition we consider again the transformation (\ref{e4}) viewed as the one relating both sets of variables at (say) $t=0$. One easily finds that 
\begin{equation}
\label{e28}
\beta=\sqrt[4]{2}(d+g).
\end{equation}
Referring to the analysis performed at the beginning we conclude that the BRST-invariance singles out the subspace spanned by the eigenvectors of the Hamiltonian. 
As we have  shown above the only state of nonvanishing norm is the vacuum, in full agreement with quartet mechanism \cite{b14}, \cite{b15}.
\par
Now the question arises what is happening if the quartic oscillator is quantized  by the method of Pais and Uhlebeck \cite{b9}.
 To this end let us note that the PU Hamiltonian is obtained from eq. (\ref{e3}) by the transformation (\ref{e4}) with $iq_2$ replaced by $q_2$ and $ip_2$ by $-p_2$. Then $q_2$
 and $p_2$ are hermitean so no modification of metric is necessary. The space of states becomes a Hilbert space while the Hamiltonian is unbounded from below. In fact, it becomes a 
difference of two commuting terms: the angular momentum  operator and radial coordinate squared \cite{b9}. This can be directly shown by considering  the canonical transformation
\begin{alignat}{2}
\label{e29}
b &=-\sqrt{2}Q_1, & \qquad \phi &=-\frac{1}{4\sqrt{2}m^2}Q_1+\frac{1}{2m}P_2, \nonumber \\
\Pi_b &=\frac{1}{4m}Q_2-\frac{1}{\sqrt{2}}P_1, & \qquad \Pi_{\phi} &=-2mQ_2,
\end{alignat}
which converts $H$, eq. (\ref{e3}), into 
\begin{equation}
\label{e30}
H=\sqrt{2}m(Q_2P_1-Q_1P_2)-\frac{1}{2}(Q_1^2+Q_2^2).
\end{equation}
The two pieces of $H$ commute. The first part has purely discrete unbounded spectrum  while the second -- purely continuous nonpositive one.
In terms of $\alpha$ and $\beta$ variables $H$ reads
\begin{align}
\label{e31}
H=&\sqrt2m(\alpha^*\beta+\beta^*\alpha+1)-2m\beta^*\beta\nonumber\\
=&(\sqrt2m(\alpha^*\beta+\beta^*\alpha+1)-m\beta^*\beta)-m\beta^*\beta.
\end{align}
On the level of these variables the Ostrogradski approach differs by the choice of the representation of the algebra (\ref{e20}).
\par Namely, one defines the new operators by 
\begin{equation}
\label{e32}
\alpha=\frac{1}{\sqrt[4]2}a_+,\quad
\beta=\sqrt[4]{2}(a_++a_-^*).
\end{equation}
They obey standard commutation rules for creation and annihilation operators 
\begin{equation}
\label{e33}
[a_\varepsilon,a^*_{\varepsilon'}]=\delta_{\varepsilon\varepsilon'}.
\end{equation}
Then the space of  of states becomes an ordinary  Hilbert space and 
\begin{equation}
\label{e34}
H=\sqrt2m(a_+^*a_+-a_-^*a_-)-\sqrt2m(a_+^*+a_-)(a_++a_-^*).
\end{equation}
One can pose the question  concerning the status of BRST-invariance condition (\ref{e27}). It is easy to see that it has no normalizable solutions. Also, the quartet
mechanism doesn't work due to the redefinition of creation and annihilation operators (this is obvious  as the bosonic sector has positive definite metric).
\par
We have considered only the simplest (1+0)-dimensional model. However, as we already mentioned above the results are valid for field-theoretical model in 1+3 dimensions.
We also restricted ourselves to the quatric case. However, the relation between BRST symmetry and  nondiagonalizability of the Hamiltonian seems to be general.
It has been noted  in Ref. {\cite{b14} that the BRST symmetry of higher-derivative theory is nilpotent provided the frequencies coincide. On the other hand, on the classical
 level the solutions to the equations with degenerate  frequencies  contain, apart from exponential also polynomial (in time) pieces. Assuming the existence of matrix  elements
of field operator  between the states belonging to the domain of Hamiltonian (which is not always the case \cite{b9},\cite{b16}) one can use the correspondence principle
 to conclude that the Hamiltonian is not diagonalizable.
\par
Let us comment finally on the interacting case. To preserve the BRST symmetry the interaction is introduced through the prepotential $U(\phi)$ \cite{b14}.
Except $U$ is chosen to be linear in $\phi$ the bosonic  and fermionic sectors do interact. For  linear $U$ we are still dealing with free case with degenerate frequencies;
the previous conclusions remain valid.
\par
For more general prepotentials the analytic solution is not available. However, one can conjecture that the Hamiltonian, when reduced to the eigenspace of ghost charge operator  
corresponding  to the eigenvalue zero, is not diagonalizable and the BRST invariance condition singles out the subspace spanned by its eigenvectors.
\par
{\bf Acknowledgments}
The discussions with C. Gonera, P. Kosi\'nski and P. Ma\'slanka are  gratefully acknowledged. 
This work  is supported  in part by  MNiSzW grant No. N202331139  (J.G.), as well as from the  earmarked subsidies MNiSzW for Young Scientists grant No. 545/199 (K.A.).

\end{document}